\documentstyle[12pt]{article}

\def\i{{\rm i}}
\def\d{{\rm d}}
\def\Lam{{\mit\Lambda}}
\def\R{\mbox{\boldmath$R$}}

\title{\marginpar{\vspace{-1in}\hspace{-1in}\small KFT U{\L} 3/93}
Non-commutative quantum dynamics\thanks{supported by KBN grant No. 2 0218 91
01}}
\author{Jakub Rembieli\'nski \& Kordian A. Smoli\'nski\\
Department of Mathematical Physics\\ University of {\L}\'od\'z\\ ul.~Pomorska
149/153\\ 90--236 {\L}\'od\'z, Poland}
\date{April 1993}

\begin{document}

\maketitle

\begin{abstract}
We described a $q$-deformation of a quantum dynamics in one
dimension. We prove that there exists only one essential
deforamtion of quantum dynamics.
\end{abstract}

\section{Introduction}

Many theoretical physicists try to construct physical models based on
non-commutative geometry \cite{connes,manin}. A large family of such models
there are deformations of standard commutative (classical and quantum)
dynamics to dynamics in $q$-deformed phase-space
\cite{arefeva,qcircle,chlopcy,dynamic,ubriaco,toy}. Other
interesting approaches in this field are the formulation of dynamics on
non-commutative configuration manifolds \cite{lagrangian} or
view on standard quantum mechanics as a form of non-commutative
geometry
\cite{dimakis} and non-commutative differential calculus
\cite{dynamic}.

In this paper we formulate unitary non-commutative $q$-dynamics on the quantum
level. Our starting point is an assumption that a possible deformation of the
standard quantum mechanics lies in change of the algebra of observables with
consequences on the level of dynamics.

We start with the well known statement, that probabilistic
interpretation of quantum mechanics causes an unitary time
evolution of physical system irrespectively of the choice of the
algebra of  observables (standard or $q$-deformed).  As a
consequence the Heisenberg equations of motion hold in each case
(in the Heisenberg picture). In the following we restrict
ourselves to the one degree of freedom systems.

The paper is organized as follows. Section~\ref{standard}.\ is
devoted to re-describing of the standard quantum mechanical
algebra of observables.  In Section~\ref{quantum}.\ we deform
the algebra of observables and find equations of motion.  In
Section~\ref{models1}.\ and \ref{models2}.\ we consider and
solve few simple models and explain why in the Aref'eva-Volovich
treatment \cite{arefeva} the unitarity of time evolution was
lost. In Section~\ref{reparametrisation}.\ is showed that we can
reparametrize the deformed theory to a ``canonical'' form.
Finally, Section~\ref{conclusions}.\ contains some conclusions
and final remarks.

\section{Observables in standard QM}\label{standard}

Construction of quantum spaces by Manin \cite{manin} as quotient of a free
algebra by two-sided ideal can be applied also to the Heisenberg algebra case.
In fact the Heisenberg algebra can be introduced as the quotient algebra
\begin{equation}\label{algebra}
{\cal H}=A(I,x.p)/J(I,x.p),
\end{equation}
where $A(I,x,p)$ is an unital associative algebra freely generated by $I$, $x$
and $p$,  while $J(I,x,p)$ is a two-sided ideal in $A$ defined by the
Heisenberg rule
\begin{equation}\label{commut}
xp=px+\i\hbar I.
\end{equation}
There is an antilinear anti-involution (star operation) in $A$ defined on
generators as below
\begin{equation}\label{hermicity}
x^*=x,\quad p^*=p.
\end{equation}
{}From the above construction it follows that this anti-involution induces in
$\cal H$ a $^*$-anti-automorphism defined again by the eqs.\
(\ref{hermicity}).

Now, according to the result of \cite{arefeva}, confirmed in
\cite{kinematics} for the relativistic case, some parameters of the
non-commutative dynamics, like inertial mass, do not commute with the
generators $x$ and $p$. This means that these parameters should be treated
themselves as generators of the algebra. Therefore it is resonable to treat
them analogously on the commutative level too. To be more concrete let us
consider a conservative system described by the Hamiltonian
\begin{equation}
H=p^2\kappa^2+V(x,\kappa,\lambda).
\label{hamiltonian}
\end{equation}
Here $\kappa$ and $\lambda$ are assumed to be additional hermitean generators
of the extended algebra $\cal H'$ satisfying the following re-ordering rules
\begin{eqnarray}
[x,p]=\i\hbar\lambda^2,\label{commut1}\\
{}[x,\lambda]=[p,\lambda]=[x,\kappa]=[p,\kappa]=[\kappa,\lambda]=0.
\label{commut2}
\end{eqnarray}

We observe that the generators $\kappa$ and $\lambda$ belong to the center of
$\cal H'$. Thus the irreducibility condition on the representation level
implies that $\lambda$ and $\kappa$ are multipliers of the identity $I$.
Consequently they can be chosen as follows
\begin{eqnarray}
\lambda&=&I,\nonumber\\
\kappa&=&\frac{1}{\sqrt{2m}}I,
\end{eqnarray}
so the extended algebra $\cal H'$ reduces to the homomorphic Heisenberg
algebra $\cal H$ defined by (\ref{algebra}) and (\ref{commut}).  Notice that
$\cal H'$ can be interpreted as a quotient of a free unital, associative and
involutive algebra $A(I,x,p,\kappa,\lambda)$ by the two-sided ideal
$J(I,x,p,\kappa,\lambda)$ defined by eqs.\ (\ref{commut1}--\ref{commut2})
i.e.\
\begin{equation}
{\cal H'}=A(I,x,p,\kappa,\lambda)/J(I,x,p,\kappa,\lambda).
\end{equation}
It is remarkable, that eqs.\ (\ref{commut1}--\ref{commut2}) are nothing but
the Bethe Ansatz for $\cal H'$.

Finally, dynamics defined by the Hamiltonian (\ref{hamiltonian}) and the
Heisenberg equations lead to the Hamilton form of the equations of motion:
\begin{eqnarray}
\dot{\lambda}&=&0,\label{hamilton1}\\
\dot{\kappa}&=&0,\\
\dot{x}&=&\frac{1}{m}p,\\
\dot{p}&=&-V'(x).\label{hamilton2}
\end{eqnarray}

\section{Observables in $q$-QM}\label{quantum}

Now, the formulation of the standard quantum mechanics by means of the algebra
$\cal H'$ suggest a natural $q$-deformation of the algebra of observables;
namely the $q$-deformed algebra ${\cal H}_{q\xi\tau\varepsilon}$ is assumed to
be a quotient algebra
\begin{equation}
{\cal H}_{q\xi\tau\varepsilon}=
A(I,x,p,K,\Lam)/J_{q\xi\tau\varepsilon}(I,x,p,K,\Lam),
\end{equation}
where the two-sided ideal $J_{q\xi\tau\varepsilon}$ is defined now by the
following Bethe Ansatz re-ordering rules
\begin{eqnarray}
xp&=&q^2px+\i\hbar q\Lam^2,\label{qcommut1}\\ x\Lam&=&\xi\Lam x,\\
p\Lam&=&\xi^{-1}\Lam p,\\ xK&=&\tau^2Kx,\\ pK&=&\varepsilon^2Kp,\\
\Lam K&=&\tau\varepsilon K\Lam,\label{qcommut2}
\end{eqnarray}
where $K$ and $\Lam$ are assumed to be invertible and
\begin{equation}
x^*=x,\quad p^*=p,\quad K^*=K,\quad \Lam^*=\Lam.
\end{equation}
A consistency of the system (\ref{qcommut1}--\ref{qcommut2}) requires
\begin{equation}
|q|=|\xi|=|\tau|=|\varepsilon|=1.
\end{equation}
The corresponding conservative Hamiltonian has the form
\begin{equation}
H=p^2K^2+V(x,K,\Lam).
\label{qhamiltonian}
\end{equation}
Now, similary to the standard case, $\Lam$ and $K$ are assumed constant in
time:
\begin{eqnarray}
\dot{\Lam}&=&\frac{\i}{\hbar}[H,\Lam]\equiv0,\label{const1}\\
\dot{K}&=&\frac{\i}{\hbar}[H,K]\equiv0,\label{const2}
\end{eqnarray}
which implies, under the requirement of the proper classical limit
(\ref{commut1}--\ref{commut2}),
\begin{eqnarray}
\varepsilon&=&1\label{eps},\\
\tau&=&\xi^{-1},\label{tau}
\end{eqnarray}
and by means of eqs.\ (\ref{eps}--\ref{tau})
\begin{eqnarray}
V(x,K,\Lam)&=&V(\xi x,\xi K,\Lam),\label{pot1}\\
V(x,K,\Lam)&=&V(\xi^2x,K,\xi\Lam).\label{pot2}
\end{eqnarray}
So, our algebra of observables ${\cal H}_{q\xi\tau\varepsilon}$ reduces to the
algebra
\begin{equation}
{\cal H}_{q\xi}=A(I,x,p,K,\Lam)/J_{q\xi}(I,x,p,K,\Lam),
\end{equation}
where now the ideal $J_{q\xi}$ is defined by rules
\begin{eqnarray}
xp&=&q^2px+\i\hbar q\Lam^2,\label{1commut1}\\ x\Lam&=&\xi\Lam x,\\
p\Lam&=&\xi^{-1}\Lam p,\\ xK&=&\xi^{-2}Kx,\\ pK&=&Kp,\\
\Lam K&=&\xi^{-1} K\Lam.\label{1commut2}
\end{eqnarray}

Finally the Heisenberg equations of motion imply its Hamiltonian form
\begin{equation}\label{xdot}
\dot{x}=\frac{\i}{\hbar}[H,x]=
\left[\frac{\i}{\hbar}(\xi^4-q^4)p^2x+q(q^2+\xi^2)p\Lam^2\right]K^2,
\end{equation}
and
\begin{eqnarray}
\lefteqn{\dot{p}=\frac{\i}{\hbar}[H,p]=
\frac{\i}{\hbar}p [{\textstyle V((\frac{q}{\xi})^2x,K,\Lam)-V(x,K,\Lam)}]+}
\nonumber\\
&& -\frac{q}{\textstyle(\frac{q}{\xi})^2-1}\frac{1}{x} [{\textstyle
V((\frac{q}{\xi})^2x,K,\Lam)-V(x,K,\Lam)}]
\Lam^2.\label{pdot}
\end{eqnarray}
Note that the last term in (\ref{pdot}) is the quantum (Gauss-Jackson)
derivative (gradient) of $V(x,K,\Lam)\Lam^2$.

If we do not assume that $\dot{x}$ is linear in $p$ (as it was {\em implicite}
done in \cite{dynamic}) we cannot reduce number of deformation parameters.

\section{Simple models (I)}\label{models1}

Now, let us consider two simple dynamical models.

\subsection{Free particle}

We choose the potential $V=0$ so
$H=\frac{1}{2\xi^2}M^{-1}p^2\Lam^{-2}$ and consequently
\begin{eqnarray}
\dot{x}&=&\frac{\i}{2\hbar}(({\textstyle\frac{q}{\xi}})^4-1)M^{-1}p^2x\Lam^{-2}
+\frac{q}{2}(({\textstyle\frac{q}{\xi}})^2+1)pM^{-1},\label{xfree}\\
\dot{p}&=&0,\label{pfree}
\end{eqnarray}
where
\begin{math}
M=\frac{\xi}{2}(K\Lam)^{-2}
\end{math} and
obeys following algebra
\begin{eqnarray}
xM&=&\xi^2Mx,\label{M1}\\ pM&=&\xi^2Mp,\\
\Lam M&=&\xi^2M\Lam.\label{M2}
\end{eqnarray}
Putting $\xi=q$ and $\Lam=I$ we obtain Aref'eva-Volovich model
\cite{arefeva}.
However the choice $\Lam=I$ leads to contradiction with the algebra
(\ref{M1}--\ref{M2}) (especially with (\ref{M2})) and causes non-unitarity of
time evolution (Heisenberg equations do not hold with this choice).

\subsection{Harmonic oscillator}

We start with the Hamiltonian:
\begin{equation}
H=\frac{1}{2\xi^2}M^{-1}p^2\Lam^{-2}+\frac{\omega^2}{2\xi^2}x^2M\Lam^{-2}.
\label{harmonic}
\end{equation}
Consequently
\begin{eqnarray}
\dot{x}&=&
-\frac{\i}{2\hbar}(({\textstyle\frac{q}{\xi}})^4-1)M^{-1}p^2x\Lam^{-2}
+\frac{q}{2}(({\textstyle\frac{q}{\xi}})^2+1)pM^{-1},\label{xharm}\\
\dot{p}&=&
\frac{\i\omega^2}{2\xi^2\hbar}(({\textstyle\frac{q}{\xi}})^4-1)px^2M\Lam^{-2}
-\frac{q\omega^2}{2\xi^2}(({\textstyle\frac{q}{\xi}})^2+1)xM.\label{pharm}
\end{eqnarray}
We get Aref'eva-Volovich model choosing $\xi=q$ and $\Lam=I$ again, but there
is no unitary time evolution for the same reason like above.

So it is evident that if we choose the inertial mass as a non-commutative
generator the algebra of observables it is necessary to append another
generator to this algebra.

\section{Reparametrisation}\label{reparametrisation}

It is easy to see that the conditions (\ref{pot1}--\ref{pot2}) are fulfiled if
we choose the potential $V$ as a function of the one variable
\begin{equation}\label{mu-potential}
V(x,K,\Lam)=V({\textstyle\frac{1}{\sqrt{\xi}}}x\mu\Lam^{-1})
\end{equation}
where we replace $K$ by a new generator
\begin{equation}\label{mu}
\mu=\sqrt{\frac{\xi}{2m}}(K\Lam)^{-1},
\end{equation}
where $m\in\R_+$ is a parameter describing value of inertial mass of particle.
Re-ordering rules for $\mu$ and the other generators are following
\begin{eqnarray}
x\mu&=&\xi\mu x,\label{mu1}\\ p\mu&=&\xi\mu p,\\
\Lam\mu&=&\xi\mu\Lam.\label{mu2}
\end{eqnarray}

Then let us reparametrize the phase-space coordinates.  We replace $x$ and $p$
by new variables $X$ and $P$ respectively:
\begin{eqnarray}
X&=&\frac{1}{\sqrt{\xi}}x\mu \Lam^{-1},\label{X}\\
P&=&\sqrt{\xi}p(\Lam\mu)^{-1}.\label{P}
\end{eqnarray}

Note that the transformation (\ref{X}--\ref{P}) is not an unitary (canonical)
transformation, so physical meaning of $x$, $p$ and $X$, $P$ are different.

With use of re-ordering rules (\ref{qcommut1}--\ref{qcommut2}) and
(\ref{mu1}--\ref{mu2}) we can find the commutation rules for the new set of
observables
\begin{eqnarray}
XP&=&\textstyle(\frac{q}{\xi})^2PX+\i\hbar(\frac{q}{\xi})I,\label{XP}\\
{}[X,\mu]&=&[X,\Lam]=[P,\mu]=[P,\Lam],\\
\Lam\mu&=&\xi\mu\Lam.\label{Commut2}
\end{eqnarray}
The energy takes the form
\begin{equation}
H=\frac{P^2}{2m}+V(X)\equiv T+V.
\label{Hamiltonian}
\end{equation}
Note that $H$ does not contain either $\mu$ or $\Lam$.

The Heisenberg equations of motion are the following
\begin{eqnarray}
\dot{X}&=&\frac{\i}{\hbar}(1-({\textstyle\frac{q}{\xi}})^4)\frac{P^2X}{2m}+
({\textstyle\frac{q}{\xi})(1+(\frac{q}{\xi}})^2)\frac{P}{2m},\label{Xdot}\\
\dot{P}&=&\frac{\i}{\hbar}P
[{\textstyle V((\frac{q}{\xi})^2X)-V(X)}]+\nonumber\\
&&-\frac{\left(\frac{q}{\xi}\right)}{\left(\frac{q}{\xi}\right)^2-1}\frac{1}{X}
[\textstyle V((\frac{q}{\xi})^2X)-V(X)],\label{Pdot}\\
\dot{\mu}&=&0,\\ \dot{\Lam}&=&0.
\end{eqnarray}
Note, that eqs.~(\ref{XP}) and (\ref{Xdot}) describing the same deformation as
in \cite{ubriaco} (under the replacement $(\frac{q}{\xi})^2\to q$).

We see that after the reparametrization (\ref{X}--\ref{P}) our algebra of
observables ${\cal H}_{q\xi}$ is a direct sum of the algebra ${\cal
H}_{q/\xi}$ given by relation (\ref{XP}) and of the real Manin's plane ${\cal
M}^2_\xi$ (generated by $\mu$ and $\Lam$).  Therefore it is evident that in
the above parametrisation under the irreducibility condition we should
restrict ourselves to the algebra generated by $X$ and $P$. It is obvious that
the ``static'' coordinates $\mu$ and $\Lam$ cannot be treated as true dynamical
variables, because they do not appear in the Hamiltonian (\ref{Hamiltonian}).

Moreover, for $q=\xi$ the theory in fact reduces to the commutative one. There
is essentially non-commutative one if we choose $\xi\neq q$; in this case
velocity is not linear but rather squared in $P$ (see eq.~(\ref{Xdot})).

An existence of classical limit $\hbar\to0$, $(q/\xi)\to1$ requires that first
terms in (\ref{Xdot}, \ref{Pdot}) have to vanish, so denoting $q/\xi={\rm
e}^{\i\theta}$ it leads to the condition on dependence of $\theta$ on $\hbar$,
namely it must be $\lim_{\hbar\to0}\frac{\d\theta(\hbar)}{\d\hbar}=0$

\section{Simple models (II)}\label{models2}

Let us turn back to the models considered in Section~\ref{models1}. Now we
show them after the reparametrisation.

\subsection{Free particle}

Now, Hamiltonian is of the form
\begin{equation}
H=\frac{P^2}{2m}
\end{equation}
and Heisenberg equations are the following
\begin{eqnarray}
\dot{X}&=&\frac{\i}{\hbar}{\textstyle(1-(\frac{q}{\xi}})^4)\frac{P^2X}{2m}+
({\textstyle\frac{q}{\xi})(1+(\frac{q}{\xi}})^2)\frac{P}{2m},\label{Xfree}\\
\dot{P}&=&0\label{Pfree}.
\end{eqnarray}
We can observe that while the momentum $P$ is all time constant, the velocity
$\dot{X}$ depends on coordinate $X$ (for $\xi\neq q$).  The explicit solution
of the eqs.~(\ref{Xfree}--\ref{Pfree}) is given in \cite{ubriaco}.

\subsection{Harmonic oscillator}

Hamiltonian is the following
\begin{equation}
H=\frac{P^2}{2m}+\frac{\omega^2X^2}{2}
\end{equation}
and Heisenberg equations of motion:
\begin{eqnarray}
\dot{X}&=&\frac{\i}{\hbar}{\textstyle(1-(\frac{q}{\xi}})^4)\frac{P^2X}{2m}+
({\textstyle\frac{q}{\xi})(1+(\frac{q}{\xi}})^2)\frac{P}{2m},
\label{Xoscill}\\
\dot{P}&=&\frac{\i\omega^2}{2\hbar}(({\textstyle\frac{q}{\xi}})^4-1)PX^2-
\frac{\omega^2}{2}{\textstyle(\frac{q}{\xi})((\frac{q}{\xi})^2+1)}X.
\label{Poscill}
\end{eqnarray}
In addition to the non-proportionality of the velocity $\dot{X}$ to the
momentum $P$, we obtain a dependence of force $\dot{P}$ on the momentum $P$.
However, the above two terms vanish in the commutative limit
($\frac{q}{\xi}\to1$).

Note that all the above equations in the commutative case have the standard
form.

\section{Conclusions}\label{conclusions}

We described the $q$-deformation of a quantum dynamics in one dimension. To
obtain the unitary time developement of observables we had to deform only the
algebra of observables leaving Heisenberg equations as equations of motion
unchanged.  We were able to reduce number of deformation parameters requiring
consistency of the algebra with Heisenberg equations of motion and finally by
the decomposition of the full algebra of observables to the direct sum of the
dynamical and internal part of this algebra. This last step is done with help
of a reparametrisation of the generators. Our final claim is that an essential
$q$-deformation of the quantum dynamics is given by the
eqs.~(\ref{XP},~\ref{Xdot},~\ref{Pdot}) with only one defermation parameter
$q/\xi$. Moreover, an essential deformation ($\xi\neq q$) leads necessarily to
quantum corrections to the velocity and force.

\section*{Acknowledgements}

We are grateful to T.~Brzezi\'nski for interesting discussions.


\end{document}